\documentclass[%
 reprint,
 amsmath,amssymb,
 aps,
 pre,
floatfix,
]{revtex4-1}

\usepackage{graphicx}
\usepackage{dcolumn}
\usepackage{bm}
\usepackage[dvipsnames]{xcolor}
\usepackage{braket}
\begin{document}

\title{First Detailed Study of the Quantum Decoherence of Entangled Gamma Photons}

\author{Julien Bordes}
\affiliation{School of Physics, Engineering and Technology, University of York, York, YO10 5DD, UK.}
\author{James R. Brown}
\affiliation{School of Physics, Engineering and Technology, University of York, York, YO10 5DD, UK.}
\email{jamie.brown@york.ac.uk}
\author{Daniel P. Watts}
\affiliation{School of Physics, Engineering and Technology, University of York, York, YO10 5DD, UK.}
\email{daniel.watts@york.ac.uk}
\author{Mikhail Bashkanov}
\affiliation{School of Physics, Engineering and Technology, University of York, York, YO10 5DD, UK.}
\author{Kieran Gibson}
\affiliation{School of Physics, Engineering and Technology, University of York, York, YO10 5DD, UK.}
\author{Ruth Newton} 
\affiliation{School of Physics, Engineering and Technology, University of York, York, YO10 5DD, UK.}
\author{Nicholas Zachariou}
\affiliation{School of Physics, Engineering and Technology, University of York, York, YO10 5DD, UK.}


\begin{abstract}
Constraints on the quantum decoherence of entangled $\gamma$ quanta at the mega-electron-volt scale, such as those produced following positron annihilation, have remained elusive for many decades. We present the first statistically and kinematically precise experimental data for triple Compton scattering of such entangled $\gamma$. An entanglement witness ($R$), relating to the enhancement of the azimuthal correlation between the final scattering planes, is obtained where one of the $\gamma$ underwent intermediate Compton scattering. The measured $R$, deconvolved from multiple scattering backgrounds, are found to exceed the classical limit for intermediate scatter angles up to $\sim$60$^{\circ}$ and diminish at larger angles. The data are consistent with predictions from a first quantum theory of entangled triple Compton scattering as well as a simple model based approach. The results are crucial to future study and utilisation of entangled mega-electron-volt $\gamma$ in fundamental physics and PET imaging.
      
\end{abstract}
\maketitle

The quantum entanglement (QE) of photonic systems in the optical or near-optical regime (electron-volt energy scale) underpins recent advances in quantum information, computing, encryption and teleportation~\cite{Slussarenko2019} and are the basis of key fundamental tests of QE~\cite{Tittel2001}. The decoherence of propagating entangled optical photons, due to the continuous monitoring of the state by the environment, is well studied theoretically and experimentally~\cite{SCHLOSSHAUER20191,decoh1,decoh2,decoh3,decoh4,decoh5,decoh6,decoh7}. However, for the mega-electron-volt scale such as positron annihilation quanta, the photons interact via different interaction processes. Our knowledge of decoherence is so poor that even the role of the leading process, Compton scattering (CS), for entangled $\gamma$ is not understood.

 Addressing this is crucial for fundamental tests of QE in new regimes of energy, for unexplored multi-partite entangled systems~\cite{Hiesmayr_2017} and societal applications, such as entangled PET imaging~\cite{Watts2021}. Entangled mega-electron-volt quanta have very different properties than optical. Detection is essentially noise free, they are more penetrating through matter (enabling deeper imaging), have wavelengths $\sim$10$^{6}$ smaller and typically small (ns) wave packet sizes. However, mega-electron-volt polarization measurement necessitates different methodologies, utilising the polarization sensitive Compton scattering (CS) process $\gamma+e^{-}\rightarrow \gamma^{\prime}+e^{-}$, described by Klein-Nishina theory~\cite{1929ZPhy...52..853K,1994okml.book.....E}. The azimuthal scatter plane ($\phi$) has a cos$^{2}\phi$ dependence relative to the $\gamma$'s linear polarization with amplitude (and a priori visibility~\cite{Hiesmayr2019}) a function of the reaction kinematics. 

The simplest source of entangled $\gamma$ quanta is positron annihilation, i.e. $e^{+}e^{-}\rightarrow 2\gamma$. The two 511~keV $\gamma$ are predicted to be produced in an entangled Bell state of linear polarization~\cite{Bohm}. 
\begin{equation}
    \ket{\Psi} = \frac{1}{\sqrt2}\: \Big( \ket{H}_{-}\ket{V}_{+}  -   \ket{V}_{-}\ket{H}_{+}\Big),
    \label{eqn:wavefunc}
\end{equation}
where $\ket{H}_{-}$ and $\ket{V}_{-}$ ($\ket{H}_{+}$ and $\ket{V}_{+}$) represent $\gamma$ with perpendicular polarizations $H$ and $V$ propagating along $-z$ ($+z$). The cross sections for double CS (DCS) of such Bell state $\gamma$ were first obtained using quantum perturbation theory~\cite{Snyder1948,Pryce1947,Thesis}. They exhibit a cos$(2\Delta\phi)$ dependence where $\Delta\phi$ is the azimuthal angle between the two Compton scatter planes~\cite{Snyder1948,Pryce1947,Thesis}. The enhancement ratio, $R$, corresponding to the ratio of yields for $\Delta\phi$=0 and 90$^{\circ}$ has a maximal value of 2.85 for symmetric polar scatter angles of 81.7$^{\circ}$. Subsequently Bohm and Aharonov~\cite{Bohm} showed this exceeded the R=1.63 predicted for DCS of a separable state viz. $\ket{H_{-}V_{+}}$ or $\ket{H_{+}V_{-}}$ and proposed measurement above this classical limit as a witness of QE~\cite{Bohm}. These results were recently confirmed in matrix~\cite{Caradonna2019} and Kraus operator formalisms~\cite{Hiesmayr_2017}).

Correlations larger than this classical limit have been observed in experiments since the 1950's, typically exploiting large NaI $\gamma$ detectors \cite{Wu1950,Langhoff1960,Kasday1971,Faraci1974,Kasday1975,Wilson1976,Bruno77,Bertolini1981}, and agree with $\Delta\phi$ correlations predicted by DCS QE theory when (analytical) estimates of experimental acceptance, resolution and backgrounds from multiple scattering (MS) within detector crystals were included. Recent progress has been aided by our implementation of the QE DCS cross section~\cite{Snyder1948,Pryce1947} into Geant4 (G4)~\cite{Agostinelli2003,Allison2016}, the leading particle transport simulation, giving QEG4~\cite{Watts2021}. This gave the first accurate simulation of entangled-$\gamma$ propagation through matter, at least for scattering orders up to DCS, and a first accurate account of detector acceptance and backgrounds. It helped establish that QE ($\Delta\phi$) information is intrinsically accessible in modern segmented $\gamma$-detector systems commensurate with those employed/planned for PET/SPECT medical scanners (e.g.~CZT~\cite{Watts2021}). Here we adopt material and detector pixel sizes matching next generation total-body PET systems (also see~\cite{Makek2019a,Makek2020,RuthThesis}). 

Despite these recent advances, further progress was limited by a lack of knowledge regarding decoherence at the mega-electron-volt scale. It was not established theoretically or experimentally what state(s) the $\gamma$ are in following the polarization analysing DCS process. In its absence a complete loss of entanglement after the first DCS was assumed in earlier key works e.g.~\cite{Kasday1975} and in QEG4~\cite{Watts2021}. Quantifying the effect of CS on entanglement requires a measurement of triple CS (TCS), in which one of the $\gamma$ undergoes an intermediate CS (ICS) before the azimuthal correlation ($\Delta\phi$) is measured. The challenge of TCS theory and measurement is finally being met, over 70 years after the first DCS experiments. TCS measurements, albeit with large bins of ICS polar angle, have recently been achieved~\cite{RuthThesis,Abdurashitov2022,Ivashkin2023,parashari2023closing}. \textcolor{black}{The first quantum theory predicting the TCS cross section for entangled $\gamma$, has also recently been derived in our group by Caradonna~\cite{Caradonna2024,Caradonna2023}}. This employs a Stokes-Mueller based matrix formalism which accounts for the propagation of the initial Bell state entanglement to the final state.

In this work, we present measurements of the enhancement ratio $R$, reflecting the azimuthal correlations between the final CS planes in a TCS process. Its behaviour is resolved for the first time with statistical precision, accurately resolved TCS kinematics and with a deconvolution of multiple scattering backgrounds. The data provide a first challenge to the recently developed quantum TCS theory as well as model based approaches.
 
The new TCS experimental data are compared to a range of simulation Ans\"atze. These are discussed below, using the nomenclature convention ($\gamma_{1,1^{^{\prime}},2,2^{\prime}}$) in Fig.~1. The QEG4-FD (full decoherence) prediction is simply the current modelling in QEG4~\cite{Watts2021}, as outlined in the introduction (See~\footnote{The current development version of QEG4 (v11.2) requires a filter on event ordering such that the polarization vectors are correctly assigned for all events.} if using the current development version (11.2) of QEG4). It predicts the azimuthal correlations in TCS where an initial Bell state $\gamma$-pair ($\gamma_{1,2}$) decoheres following the first DCS into two separable photon states ($\gamma_{1^{\prime},2^{\prime}}$). It therefore represents the TCS classical limit where the intermediate DCS follows the established QE theory~\cite{Snyder1948,Pryce1947}. We note the $R$ are within $\leq$5\% of those obtained with $\gamma_{1^{},2^{}}$ initially in a (hypothetical) separable state, i.e. $\ket{H_{-}V_{+}}$ indicating the classical limit appears robust to assumptions about the intermediate DCS. The G4-unpol simulation employs unpolarised and separable photons ($\gamma_{1,2}$) giving $R$=1~\cite{Bohm} and is of utility to constrain experimental biases (see results).

Crucial to the analysis of the data is implementation of theories for (fully) entangled TCS in simulation. The 3CG4-Caradonna simulation incorporates the first quantum theoretical calculation of this process~\cite{Caradonna2024,Caradonna2023}. A further simulation (QEG4-ENT) provides predictions for TCS where the initial entanglement (between $\gamma_{1}$ and $\gamma_{2}$) is maintained between $\gamma_{1}$ and $\gamma_{2}^{\prime}$, with the ICS only changing the direction and energy of $\gamma_{2}^{\prime}$. The DCS of $\gamma_{1}$ and $\gamma_{2}^{\prime}$ are therefore as expected from the Bell state (Eqn. 1), with cross section for non-identical $\gamma$ energies obtained using the ``partial polarization" Ansatz of Snyder et al.~\cite{Snyder1948} (see supplemental materials).

\begin{figure}[htp]
  \centering
 \includegraphics[width=1.0\columnwidth]{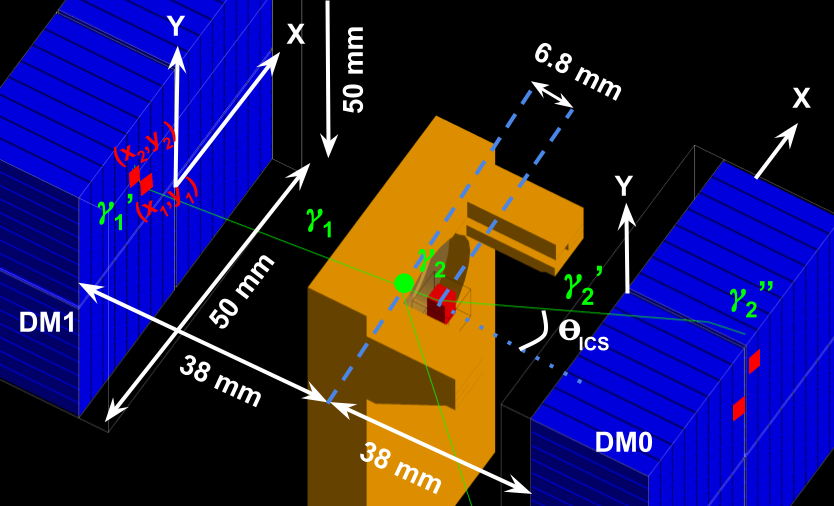}
    \caption {G4 visualisation showing source location (green circle), holder (orange), SCD (red box), LYSO pixels of the DM0/DM1 detectors (blue). Detector windows/enclosures shown by white lines, PCBs omitted for clarity. Green lines show particle tracks from a single TCS event. The $\gamma$ referencing ($\gamma_{1,1^{\prime},2,2^{\prime},2^{\prime\prime}}$), is identified next to the tracks. Red pixels show the energy deposits in DM0/1. 
    }
    \label{fig:G4setup}
\end{figure}
A G4 visualisation of the apparatus is shown in Fig.\ref{fig:G4setup}. Two segmented detector modules (DM) each comprising 256 3x3x20 mm\textsuperscript{3} LYSO scintillator crystals coupled to Ketek PM3325 silicon photomultiplier arrays were positioned either side of a $\sim$1.8~MBq \textsuperscript{22}Na positron source, with entrances 38~mm from the source. An additional 3x3x5~mm\textsuperscript{3} LYSO \emph{scatter} detector (SCD) was placed between the source and DM0, with centre 6.8~mm from the source and longest side oriented vertically. Data acquisition used PETsys ASICs \cite{DiFrancesco2016}. The CS polar angle in the SCD ($\theta_{ICS}$) was calculated from the deposited energy. A data set was also obtained with DM0 rotated $28^\circ$ about the source position. 

The particle tracks from a typical TCS event are shown in Fig.~\ref{fig:G4setup} (green lines), comprising a CS in the SCD coincident with CS candidate events  in DM0/DM1 (viz. double hits corresponding to the CS site and detection of the scattered $\gamma$). The experimental trigger was a 15~ns coincidence between SCD and DM0/1 ($\pm$2~ns achieved offline). Calibration using laboratory sources and intrinsic radioactivity in the LYSO provided six points (31.4--1274.5~keV), with calibration from detector pulse height to energy obtained from a 4\textsuperscript{th} order polynomial fit. The SCD energy resolution was $12.2\%$ (FWHM) at 511 keV and detection threshold was $\sim7$~keV. Gain drifts in SCD and DM0/1 ($\sim$5\%) were corrected using the 511~keV photopeak. The energy resolution for summed double hits (SDH) (see Fig.~\ref{fig:G4setup}) were $12.4\%$ ($14.3\%$) FWHM for DM0 (DM1). The full apparatus and source was reproduced in the QEG4 simulation. Predicted energy deposits were smeared according to the resolutions above. SDH peak centroids in data and simulation agreed within $\sim$2\%, with discrepancy attributable to differences in SiPM light collection for crystal entry points of CS $\gamma$ (through side faces) cf. front face entry of calibration $\gamma$. 
 
The experimental data and the simulated data were analysed with the same analysis code. The CS polar angles in each head were calculated assuming the larger of the two energy deposits corresponded to the location and energy of the CS. For discussion see~\cite{Watts2021}.
\begin{figure}[ht]
  \centering
    \includegraphics[width=\columnwidth]  {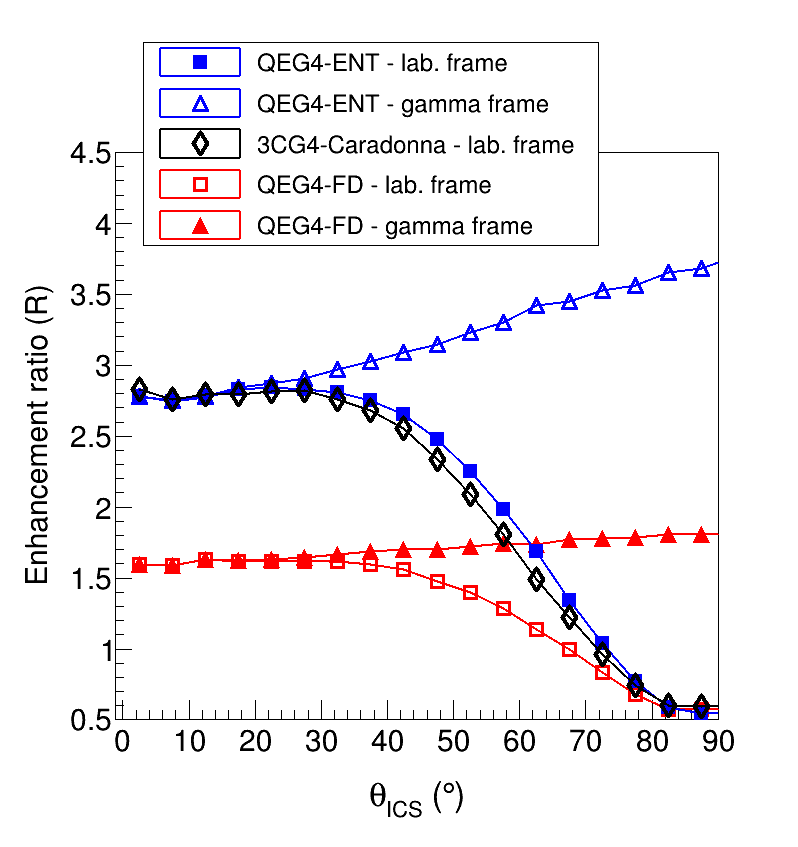}
      \caption{$R$ extracted from a ``perfect detector" (see text) for $\theta_{1,2'}\in{80^{\circ}-84^{\circ}}$.
    Red data points show QEG4-FD predictions in the laboratory frame (triangles) and photon frame (open squares). Corresponding for QEG4-ENT are shown by blue markers. The 3CG4-Caradonna theory is shown (in the laboratory frame only) by the open black diamond markers. }
    \label{fig:water}
\end{figure}
\begin{figure}[ht]
  \centering
    \includegraphics[width=\columnwidth]{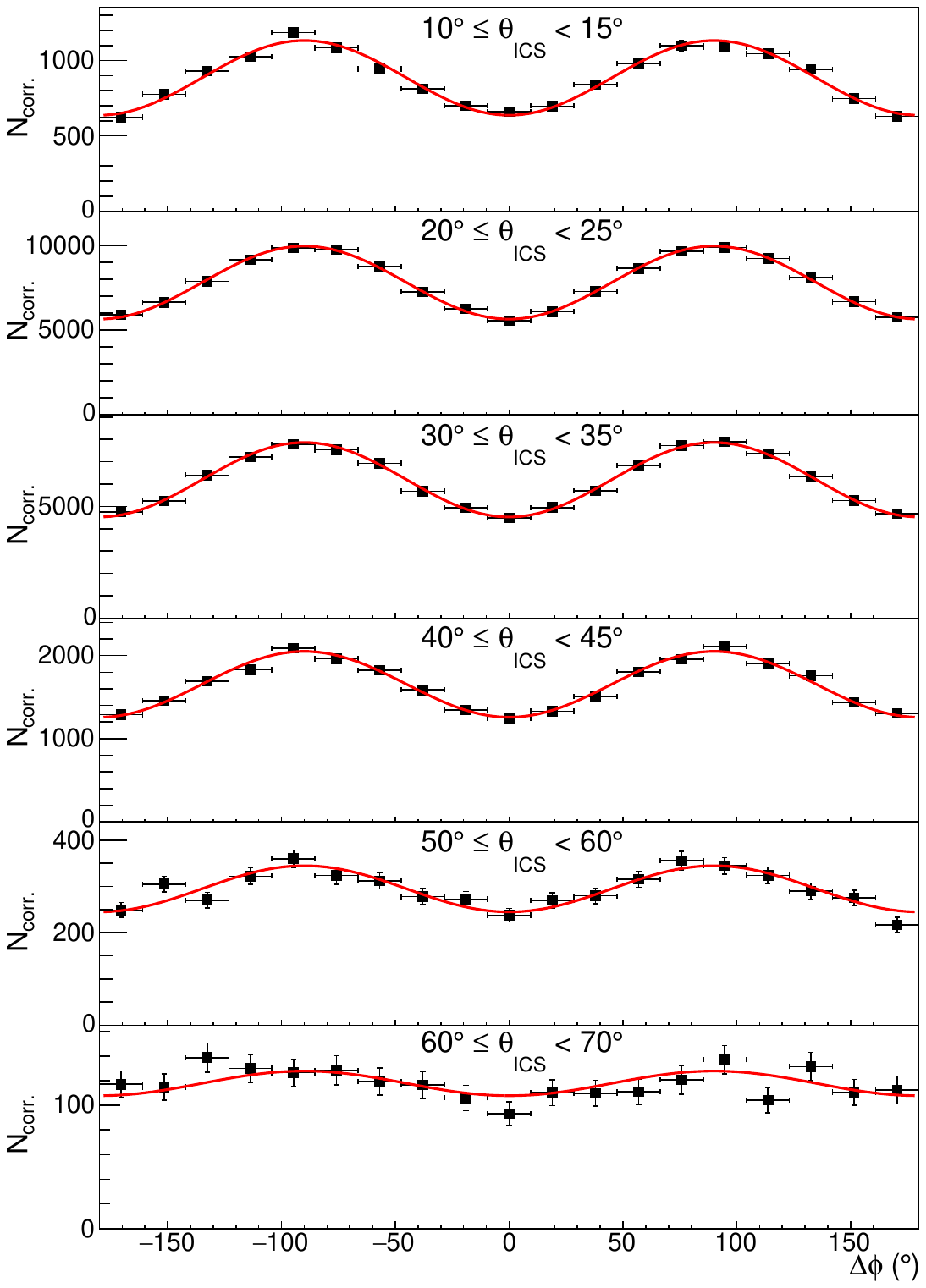}
    \caption{Measured $\Delta\phi$ distributions for events with $\theta_{1,2'}\in{72^{\circ}-92^{\circ}}$ for a range of $\theta_{ICS}$ bins (indicated in each panel). Red lines show the fits to extract $R$ \textcolor{black}{(see text)}. 
    }
    \label{fig:dPhi_theta}
\end{figure}
\begin{figure}[htp]
  \centering
    \includegraphics[width=\columnwidth]{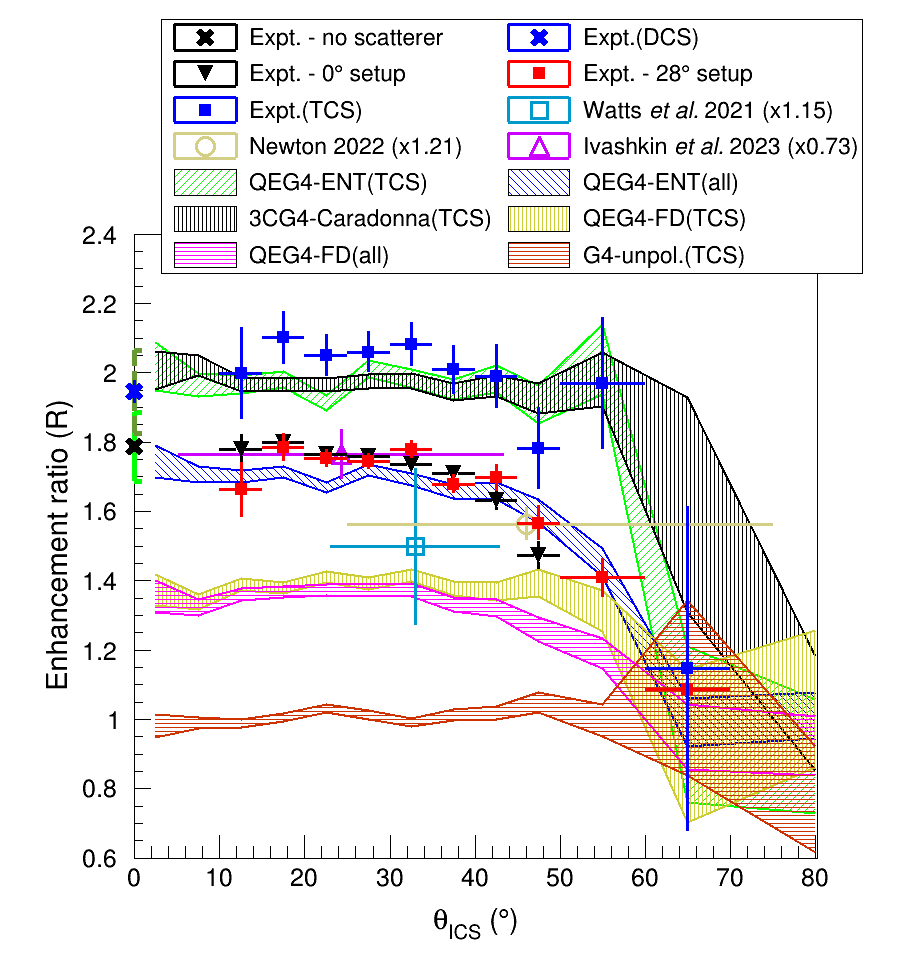}
    \caption{$R$ versus $\theta_{ICS}$ for $\theta_{1,2'}\in{72^{\circ}-92^{\circ}}$. Black (red) data points for back-to-back (rotated) detector configurations and the average of these data deconvolved of MS backgrounds (blue squares). Datum at $\theta_{ICS}$=0 (black cross) from back-to-back apparatus without SCD (green bar shows systematic error - see text). Blue cross shows same deconvolved of MS. Turquoise, beige, purple data points show previous data (normalised - see legend) from CZT~\cite{Watts2021}, LYSO~\cite{RuthThesis} and NaI~\cite{Ivashkin2023} (x-errors show $\theta_{ICS}$ acceptance). Pink (blue) bands show QEG4-FD (QEG4-ENT) predictions for all events in the acceptance. Yellow (green) bands show corresponding for TCS with black band for 3CG4-Caradonna. Red band shows the G4-unpol. predictions.}
    \label{fig:R_vs_theta_polComp}
\end{figure}
The $\phi$ measured at the detector faces were calculated using the location of the two pixel-hits in each DM array, $x_{1,2}$ and $y_{1,2}$ (Fig.~\ref{fig:G4setup}) with 
$\phi=\arctan\left(\frac{y_{2}-y_{1}}{x_{2}-x_{1}}\right)$
 and resolution $\sim$1.5$^{\circ}$--$12^\circ$ dependent on pixel separation (adjacent pixel hits were excluded). The relative azimuthal scattering angle is $\Delta\phi=\phi_1-\phi_2'$. The $\Delta\phi$ correlation amplitude is extracted using the function Acos(2$\Delta\phi$)+B. The enhancement ($R$), the ratio of counts between $\Delta\phi=90^\circ$ to $0^\circ$, was $R=(B-A)/(B+A)$. 
The effect of detector acceptance was corrected using event-mixing of uncorrelated events in the data (see supplemental materials) but had negligible ($\leq$2\%) influence on the extracted $R$.

\label{sec:res}

 Before presenting the \textcolor{black}{predictions} within the detector acceptance it is informative to obtain predictions for a ``perfect" detector. This was a (simulated) isotropic water sphere, with the source emitting $\gamma$ along the z axis from its centre. We use the exact hit locations and energy deposits from simulation and retain only TCS event. The extracted $R$ for $\theta_{1,2^{\prime}}\in{80^{\circ}-84^{\circ}}$ \textcolor{black}{(a region around the maximum enhancement at symmetric $81.7^{\circ}$ scatter angles~\cite{Snyder1948})} are shown (Fig.~\ref{fig:water}). For QEG4-FD the open red square markers show the $R$ extracted in a fixed laboratory frame ($\hat{z}$ in beam direction with fixed $\hat{x}$, $\hat{y}$). For $\theta_{ICS}\rightarrow0^{\circ}$, $R$ converges near to the expected~\cite{Bohm} value ($R=1.63$) for a separable state in DCS. The red triangular data points show the QEG4-FD $R$ values extracted in the $\gamma$ frame, where the $\phi$ angle of $\gamma_{2}^{\prime}$ is determined with respect to its polarization (known for each simulated event). In this frame $R$ rises with $\theta_{ICS}$, reflecting increasing CS analysing power with reducing energy of $\gamma_{2}^{\prime}$. The QEG4-ENT model (blue markers) shows the (expected)~\cite{Bohm} $R$ value of $\sim$2.7 at small $\theta_{ICS}$. For larger $\theta_{ICS}$, similar general features as QEG4-FD are evident between the two frames. The 3CG4-Caradonna theory in the laboratory frame (open black diamond markers) has similar features as QEG4-ENT, with almost exact correspondence up to $\theta_{ICS}\sim$30$^{\circ}$ and for regions of negative $R$ at large angles. Between these regions the agreement is still typically within 5\%. As the intermediate $\gamma$ polarization is not single valued in the TCS theory, predictions in the $\gamma$ frame cannot be simply defined.

Examples of the experimental $\Delta\phi$ distributions and $R$ fits (in the laboratory frame) for different $\theta_{ICS}$ bins are shown in Fig.~\ref{fig:dPhi_theta}. Systematic errors in $R$ are estimated to be $\sim$5\%, derived from asymmetries in $\pm\Delta\phi$ (1.4\%), variation of DM0/1/SCD energy calibrations (2\%), uncertainties in DM0/1 in-array crystal locations (1\%) and DM0/1/SCD locations (2\%). The extracted $R$ are presented as a function of $\theta_{ICS}$ in Fig.~\ref{fig:R_vs_theta_polComp} with black (red) data points for back-to-back (DM0 rotated through 28$^\circ$), showing consistency to $\leq$8\% where they overlap. Below $\theta_{ICS}\sim35^{\circ}$ $R$ shows a modest decrease ($\sim$6\%) with increasing $\theta_{ICS}$. For larger angles, $R$ diminishes more rapidly until $\sim$ $70^{\circ}$, approaching $R$=1. The datum at $\theta_{ICS}=0^\circ$ (black cross) is obtained without SCD but with DM1 acceptance cuts to match the hit distribution of SCD coincidence events, (estimated systematic in this procedure shown by the green error bar).

 Previous TCS data~\cite{Watts2021,RuthThesis, Abdurashitov2022,Ivashkin2023} are also shown in Fig.~\ref{fig:R_vs_theta_polComp} (also see unpublished results~\cite{parashari2023closing}). Since the measured $R$ is apparatus dependent, for clarity of presentation the other datasets are normalised to the $R$ of the current data at $\theta_{ICS}\sim 0^{\circ}$ (normalisations in the legend). Although this procedure is informative to compare data quality and observe trends, robust comparison with earlier data would require simulation of their specific MS contributions, acceptances and resolutions which influence the $\theta_{ICS}$ dependencies. With this caveat, the trends in our data shows broad consistency with previous data when normalised to a common $R$ at $\theta_{ICS}=0$. It is clear that the new data resolve the behaviour of the $R$ entanglement witness in detail for the first time.

The data are also compared with theory and model based predictions (bands) in Fig.~\ref{fig:R_vs_theta_polComp}. The band widths represent the statistical accuracy (each derived from $\sim$3 trillion simulated annihilations). Where appropriate, the predictions are shown with bracketed postscripts ``all" for the unrestricted event sample and ``TCS" for isolated TCS events. The G4-unpol(TCS) prediction is flat ($R\sim$1), showing no significant ``false" $R$ from detector acceptances. The QEG4-FD predictions, providing the classical limit from a (decohered) separable state, are shown by the purple (yellow) bands for all (TCS). The two scenarios give consistent $R$ except at large $\theta_{ICS}$, where the unrestricted data has increased contributions from MS backgrounds in which  a $\gamma$ has $>$1 interaction in a detector pixel. The data exceed the classical limit for most of the $\theta_{ICS}$ range, providing a witness of entanglement.

 Deeper insights can be obtained by comparison with predictions incorporating QE in TCS. The QEG4-ENT(all) predictions (blue band) reproduce the measured $R$ within $\sim$10\%. However, a clearer interpretation is possible from comparison with data where the MS backgrounds have been deconvolved to determine $R$ from a ``pure" TCS process. The deconvolved TCS (DCS) experimental data are shown by the blue square (blue cross) markers (TCS from an average of the 0$^{\circ}$ and 28$^{\circ}$ datasets). The TCS deconvolution function was $R_{expt}^{TCS} = \{\{R_{expt}^{all}-R_{sim}^{all}\}/f_{sim}^{TCS}\}+R_{sim}^{TCS}$, with enhancements ($R_{sim}^{TCS}$, $R_{sim}^{all}$) and TCS fraction ($f_{sim}^{TCS}$), along with their statistical errors, taken from the QEG4-ENT(all/TCS) predictions in each data bin. The 3CG4-Caradonna theory reproduces this TCS data over the entire kinematic range, a first experimental confirmation of the validity of any QE TCS theory. The QEG4-ENT(TCS) model also reproduces the data. The data (and both predictions) are consistent with the expected~\cite{Caradonna2024,Caradonna2023} convergence of TCS $R$ to the DCS value as $\theta_{ICS}\rightarrow0$. 

Future work will further challenge the TCS theory with new measurements having larger kinematic acceptances, including larger ranges of polar and azimuthal intermediate scattering angles. Work to extend the studies to 4CS are underway, which will improve constraints on the MS backgrounds in our (and any currently feasible) measurements of TCS. As suggested in~\cite{Watts2021}, the sensitivity to scatter backgrounds in QE-PET imaging should be re-assessed now quality TCS data and theory is available (the sensitivity to random backgrounds is unaffected).

In summary, we present the first statistically and kinematically precise measurements of an $R$ entanglement witness for the triple Compton scattering of entangled $\gamma$ photons, the most fundamental reaction to constrain entanglement decoherence at the mega-electron-volt scale. Data are obtained for intermediate Compton scattering angles $\theta_{ICS}=0-70^{\circ}$. Multiple scattering backgrounds are deconvolved from the experimental data. The resulting $R$ is flat with value $\sim$2.1 for $\theta_{ICS}=0-60^{\circ}$, clearly exceeding the classical limit (for separable $\gamma$) established within our detector apparatus ($R\sim$1.4). The $R$ data are well described over the full measured range by the entangled 3-Compton theory of Caradonna~\cite{Caradonna2024,Caradonna2023} which takes into account potential quantum-decoherence effects. The results provide a step forward in our understanding of the fundamental nature of quantum entanglement and its decoherence at the mega-electron-volt scale, crucial to a new generation of fundamental tests as well as societal applications, such as quantum entangled PET imaging. 

 \begin{acknowledgments}
We thank D. Jenkins for insights and comments and K. See for undergraduate project work. The simulation work was undertaken on the Viking Cluster, a high performance computing facility provided by the University of York. We are grateful for computational support from the University of York High Performance Computing service, Viking and the Research Computing team. The authors gratefully acknowledge the support of UKRI QTFP Grant No. ST/W006383/1 and Science and Technology Facilities Council (STFC) Grant No. ST/V001035/1.
\end{acknowledgments}
 
%


\begin{thebibliography}{43}%
\makeatletter
\providecommand \@ifxundefined [1]{%
 \@ifx{#1\undefined}
}%
\providecommand \@ifnum [1]{%
 \ifnum #1\expandafter \@firstoftwo
 \else \expandafter \@secondoftwo
 \fi
}%
\providecommand \@ifx [1]{%
 \ifx #1\expandafter \@firstoftwo
 \else \expandafter \@secondoftwo
 \fi
}%
\providecommand \natexlab [1]{#1}%
\providecommand \enquote  [1]{``#1''}%
\providecommand \bibnamefont  [1]{#1}%
\providecommand \bibfnamefont [1]{#1}%
\providecommand \citenamefont [1]{#1}%
\providecommand \href@noop [0]{\@secondoftwo}%
\providecommand \href [0]{\begingroup \@sanitize@url \@href}%
\providecommand \@href[1]{\@@startlink{#1}\@@href}%
\providecommand \@@href[1]{\endgroup#1\@@endlink}%
\providecommand \@sanitize@url [0]{\catcode `\\12\catcode `\$12\catcode `\&12\catcode `\#12\catcode `\^12\catcode `\_12\catcode `\%12\relax}%
\providecommand \@@startlink[1]{}%
\providecommand \@@endlink[0]{}%
\providecommand \url  [0]{\begingroup\@sanitize@url \@url }%
\providecommand \@url [1]{\endgroup\@href {#1}{\urlprefix }}%
\providecommand \urlprefix  [0]{URL }%
\providecommand \Eprint [0]{\href }%
\providecommand \doibase [0]{http://dx.doi.org/}%
\providecommand \selectlanguage [0]{\@gobble}%
\providecommand \bibinfo  [0]{\@secondoftwo}%
\providecommand \bibfield  [0]{\@secondoftwo}%
\providecommand \translation [1]{[#1]}%
\providecommand \BibitemOpen [0]{}%
\providecommand \bibitemStop [0]{}%
\providecommand \bibitemNoStop [0]{.\EOS\space}%
\providecommand \EOS [0]{\spacefactor3000\relax}%
\providecommand \BibitemShut  [1]{\csname bibitem#1\endcsname}%
\let\auto@bib@innerbib\@empty
\bibitem [{\citenamefont {Slussarenko}\ and\ \citenamefont {Pryde}(2019)}]{Slussarenko2019}%
  \BibitemOpen
  \bibfield  {author} {\bibinfo {author} {\bibfnamefont {S.}~\bibnamefont {Slussarenko}}\ and\ \bibinfo {author} {\bibfnamefont {G.~J.}\ \bibnamefont {Pryde}},\ }\href {htttps://doi.org/10.1063/1.5115814} {\bibfield  {journal} {\bibinfo  {journal} {Appl. Phys. Rev.}\ }\textbf {\bibinfo {volume} {6}},\ \bibinfo {pages} {041303} (\bibinfo {year} {2019})}\BibitemShut {NoStop}%
\bibitem [{\citenamefont {Tittel}\ and\ \citenamefont {Weihs}(2001)}]{Tittel2001}%
  \BibitemOpen
  \bibfield  {author} {\bibinfo {author} {\bibfnamefont {W.}~\bibnamefont {Tittel}}\ and\ \bibinfo {author} {\bibfnamefont {G.}~\bibnamefont {Weihs}},\ }\href {\doibase 10.26421/qic1.2-2} {\bibfield  {journal} {\bibinfo  {journal} {Quantum Inf. Comput.}\ }\textbf {\bibinfo {volume} {1}},\ \bibinfo {pages} {3} (\bibinfo {year} {2001})}\BibitemShut {NoStop}%
\bibitem [{\citenamefont {Schlosshauer}(2019)}]{SCHLOSSHAUER20191}%
  \BibitemOpen
  \bibfield  {author} {\bibinfo {author} {\bibfnamefont {M.}~\bibnamefont {Schlosshauer}},\ }\href {\doibase https://doi.org/10.1016/j.physrep.2019.10.001} {\bibfield  {journal} {\bibinfo  {journal} {Phys. Rep.}\ }\textbf {\bibinfo {volume} {831}},\ \bibinfo {pages} {1} (\bibinfo {year} {2019})}\BibitemShut {NoStop}%
\bibitem [{\citenamefont {Zurek}(1981)}]{decoh1}%
  \BibitemOpen
  \bibfield  {author} {\bibinfo {author} {\bibfnamefont {W.~H.}\ \bibnamefont {Zurek}},\ }\href@noop {} {\bibfield  {journal} {\bibinfo  {journal} {Phys. Rev. D}\ }\textbf {\bibinfo {volume} {24}},\ \bibinfo {pages} {1516} (\bibinfo {year} {1981})}\BibitemShut {NoStop}%
\bibitem [{\citenamefont {Zurek}(1982)}]{decoh2}%
  \BibitemOpen
  \bibfield  {author} {\bibinfo {author} {\bibfnamefont {W.~H.}\ \bibnamefont {Zurek}},\ }\href@noop {} {\bibfield  {journal} {\bibinfo  {journal} {Phys. Rev. D}\ }\textbf {\bibinfo {volume} {26}},\ \bibinfo {pages} {1862} (\bibinfo {year} {1982})}\BibitemShut {NoStop}%
\bibitem [{\citenamefont {Paz}\ and\ \citenamefont {Zurek}(2002)}]{decoh3}%
  \BibitemOpen
  \bibfield  {author} {\bibinfo {author} {\bibfnamefont {J.~P.}\ \bibnamefont {Paz}}\ and\ \bibinfo {author} {\bibfnamefont {W.~H.}\ \bibnamefont {Zurek}},\ }in\ \href@noop {} {\emph {\bibinfo {booktitle} {Fundamentals of quantum information: quantum computation, communication, decoherence and all that}}}\ (\bibinfo  {publisher} {Springer},\ \bibinfo {year} {2002})\ pp.\ \bibinfo {pages} {77--148}\BibitemShut {NoStop}%
\bibitem [{\citenamefont {Zurek}(2003)}]{decoh4}%
  \BibitemOpen
  \bibfield  {author} {\bibinfo {author} {\bibfnamefont {W.~H.}\ \bibnamefont {Zurek}},\ }\href@noop {} {\bibfield  {journal} {\bibinfo  {journal} {Rev. Mod. Phys.}\ }\textbf {\bibinfo {volume} {75}},\ \bibinfo {pages} {715} (\bibinfo {year} {2003})}\BibitemShut {NoStop}%
\bibitem [{\citenamefont {Schlosshauer}(2005)}]{decoh5}%
  \BibitemOpen
  \bibfield  {author} {\bibinfo {author} {\bibfnamefont {M.}~\bibnamefont {Schlosshauer}},\ }\href@noop {} {\bibfield  {journal} {\bibinfo  {journal} {Reviews of Modern physics}\ }\textbf {\bibinfo {volume} {76}},\ \bibinfo {pages} {1267} (\bibinfo {year} {2005})}\BibitemShut {NoStop}%
\bibitem [{\citenamefont {Bacciagaluppi}(2020)}]{decoh6}%
  \BibitemOpen
  \bibfield  {author} {\bibinfo {author} {\bibfnamefont {G.}~\bibnamefont {Bacciagaluppi}},\ }in\ \href@noop {} {\emph {\bibinfo {booktitle} {The {Stanford} Encyclopedia of Philosophy}}},\ \bibinfo {editor} {edited by\ \bibinfo {editor} {\bibfnamefont {E.~N.}\ \bibnamefont {Zalta}}}\ (\bibinfo  {publisher} {Metaphysics Research Lab, Stanford University},\ \bibinfo {year} {2020})\ \bibinfo {edition} {{F}all 2020}\ ed.\BibitemShut {Stop}%
\bibitem [{\citenamefont {Joos}\ \emph {et~al.}(2013)\citenamefont {Joos}, \citenamefont {Zeh}, \citenamefont {Kiefer}, \citenamefont {Giulini}, \citenamefont {Kupsch},\ and\ \citenamefont {Stamatescu}}]{decoh7}%
  \BibitemOpen
  \bibfield  {author} {\bibinfo {author} {\bibfnamefont {E.}~\bibnamefont {Joos}}, \bibinfo {author} {\bibfnamefont {H.~D.}\ \bibnamefont {Zeh}}, \bibinfo {author} {\bibfnamefont {C.}~\bibnamefont {Kiefer}}, \bibinfo {author} {\bibfnamefont {D.~J.}\ \bibnamefont {Giulini}}, \bibinfo {author} {\bibfnamefont {J.}~\bibnamefont {Kupsch}}, \ and\ \bibinfo {author} {\bibfnamefont {I.-O.}\ \bibnamefont {Stamatescu}},\ }\href@noop {} {\emph {\bibinfo {title} {Decoherence and the appearance of a classical world in quantum theory}}}\ (\bibinfo  {publisher} {Springer Science \& Business Media},\ \bibinfo {year} {2013})\BibitemShut {NoStop}%
\bibitem [{\citenamefont {Hiesmayr}\ and\ \citenamefont {Moskal}(2017)}]{Hiesmayr_2017}%
  \BibitemOpen
  \bibfield  {author} {\bibinfo {author} {\bibfnamefont {B.~C.}\ \bibnamefont {Hiesmayr}}\ and\ \bibinfo {author} {\bibfnamefont {P.}~\bibnamefont {Moskal}},\ }\href {http://www.nature.com/articles/s41598-017-15356-y} {\bibfield  {journal} {\bibinfo  {journal} {Sci. Rep.}\ }\textbf {\bibinfo {volume} {7}},\ \bibinfo {pages} {5} (\bibinfo {year} {2017})}\BibitemShut {NoStop}%
\bibitem [{\citenamefont {Watts}\ \emph {et~al.}(2021)\citenamefont {Watts}, \citenamefont {Bordes}, \citenamefont {Brown}, \citenamefont {Cherlin}, \citenamefont {Newton}, \citenamefont {Allison}, \citenamefont {Bashkanov}, \citenamefont {Efthimiou},\ and\ \citenamefont {Zachariou}}]{Watts2021}%
  \BibitemOpen
  \bibfield  {author} {\bibinfo {author} {\bibfnamefont {D.~P.}\ \bibnamefont {Watts}}, \bibinfo {author} {\bibfnamefont {J.}~\bibnamefont {Bordes}}, \bibinfo {author} {\bibfnamefont {J.~R.}\ \bibnamefont {Brown}}, \bibinfo {author} {\bibfnamefont {A.}~\bibnamefont {Cherlin}}, \bibinfo {author} {\bibfnamefont {R.}~\bibnamefont {Newton}}, \bibinfo {author} {\bibfnamefont {J.}~\bibnamefont {Allison}}, \bibinfo {author} {\bibfnamefont {M.}~\bibnamefont {Bashkanov}}, \bibinfo {author} {\bibfnamefont {N.}~\bibnamefont {Efthimiou}}, \ and\ \bibinfo {author} {\bibfnamefont {N.~A.}\ \bibnamefont {Zachariou}},\ }\href {\doibase 10.1038/s41467-021-22907-5} {\bibfield  {journal} {\bibinfo  {journal} {Nat. Commun.}\ }\textbf {\bibinfo {volume} {12}},\ \bibinfo {pages} {2646} (\bibinfo {year} {2021})}\BibitemShut {NoStop}%
\bibitem [{\citenamefont {{Klein}}\ and\ \citenamefont {{Nishina}}(1929)}]{1929ZPhy...52..853K}%
  \BibitemOpen
  \bibfield  {author} {\bibinfo {author} {\bibfnamefont {O.}~\bibnamefont {{Klein}}}\ and\ \bibinfo {author} {\bibfnamefont {T.}~\bibnamefont {{Nishina}}},\ }\href {\doibase 10.1007/BF01366453} {\bibfield  {journal} {\bibinfo  {journal} {Zeitschrift fur Physik}\ }\textbf {\bibinfo {volume} {52}},\ \bibinfo {pages} {853} (\bibinfo {year} {1929})}\BibitemShut {NoStop}%
\bibitem [{\citenamefont {{Ekspong}}(1994)}]{1994okml.book.....E}%
  \BibitemOpen
  \bibfield  {author} {\bibinfo {author} {\bibfnamefont {G.}~\bibnamefont {{Ekspong}}},\ }\href {\doibase 10.1142/2074} {\emph {\bibinfo {title} {{The Oskar Klein Memorial Lectures. Vol. 2: Lectures by Hans A. Bethe and Alan H. Guth, with translated reprints by Oskar Klein.}}}}\ (\bibinfo {year} {1994})\BibitemShut {NoStop}%
\bibitem [{\citenamefont {Hiesmayr}\ and\ \citenamefont {Moskal}(2019)}]{Hiesmayr2019}%
  \BibitemOpen
  \bibfield  {author} {\bibinfo {author} {\bibfnamefont {B.~C.}\ \bibnamefont {Hiesmayr}}\ and\ \bibinfo {author} {\bibfnamefont {P.}~\bibnamefont {Moskal}},\ }\href {\doibase 10.1038/s41598-019-44570-z} {\bibfield  {journal} {\bibinfo  {journal} {Sci. Rep.}\ }\textbf {\bibinfo {volume} {9}},\ \bibinfo {pages} {1} (\bibinfo {year} {2019})}\BibitemShut {NoStop}%
\bibitem [{\citenamefont {Bohm}\ and\ \citenamefont {Aharonov}(1957)}]{Bohm}%
  \BibitemOpen
  \bibfield  {author} {\bibinfo {author} {\bibfnamefont {D.}~\bibnamefont {Bohm}}\ and\ \bibinfo {author} {\bibfnamefont {Y.}~\bibnamefont {Aharonov}},\ }\href {\doibase 10.1103/PhysRev.108.1070} {\bibfield  {journal} {\bibinfo  {journal} {Phys. Rev.}\ }\textbf {\bibinfo {volume} {108}},\ \bibinfo {pages} {1070} (\bibinfo {year} {1957})}\BibitemShut {NoStop}%
\bibitem [{\citenamefont {Snyder}\ \emph {et~al.}(1948)\citenamefont {Snyder}, \citenamefont {Pasternack},\ and\ \citenamefont {Hornbostel}}]{Snyder1948}%
  \BibitemOpen
  \bibfield  {author} {\bibinfo {author} {\bibfnamefont {H.~S.}\ \bibnamefont {Snyder}}, \bibinfo {author} {\bibfnamefont {S.}~\bibnamefont {Pasternack}}, \ and\ \bibinfo {author} {\bibfnamefont {J.}~\bibnamefont {Hornbostel}},\ }\href@noop {} {\bibfield  {journal} {\bibinfo  {journal} {Phys. Rev.}\ }\textbf {\bibinfo {volume} {73}},\ \bibinfo {pages} {440} (\bibinfo {year} {1948})}\BibitemShut {NoStop}%
\bibitem [{\citenamefont {Pryce}\ and\ \citenamefont {Ward}(1947)}]{Pryce1947}%
  \BibitemOpen
  \bibfield  {author} {\bibinfo {author} {\bibfnamefont {M.~H.~L.}\ \bibnamefont {Pryce}}\ and\ \bibinfo {author} {\bibfnamefont {J.~C.}\ \bibnamefont {Ward}},\ }\href {\doibase 10.1038/160435a0} {\bibfield  {journal} {\bibinfo  {journal} {Nature}\ }\textbf {\bibinfo {volume} {160}},\ \bibinfo {pages} {435} (\bibinfo {year} {1947})}\BibitemShut {NoStop}%
\bibitem [{\citenamefont {Ward}(1949)}]{Thesis}%
  \BibitemOpen
  \bibfield  {author} {\bibinfo {author} {\bibfnamefont {J.~C.}\ \bibnamefont {Ward}},\ }\href@noop {} {\bibfield  {journal} {\bibinfo  {journal} {PhD. thesis, University of Oxford, UK}\ } (\bibinfo {year} {1949})}\BibitemShut {NoStop}%
\bibitem [{\citenamefont {Caradonna}\ \emph {et~al.}(2019)\citenamefont {Caradonna}, \citenamefont {Reutens}, \citenamefont {Takahashi}, \citenamefont {Takeda},\ and\ \citenamefont {Vegh}}]{Caradonna2019}%
  \BibitemOpen
  \bibfield  {author} {\bibinfo {author} {\bibfnamefont {P.}~\bibnamefont {Caradonna}}, \bibinfo {author} {\bibfnamefont {D.}~\bibnamefont {Reutens}}, \bibinfo {author} {\bibfnamefont {T.}~\bibnamefont {Takahashi}}, \bibinfo {author} {\bibfnamefont {S.}~\bibnamefont {Takeda}}, \ and\ \bibinfo {author} {\bibfnamefont {V.}~\bibnamefont {Vegh}},\ }\href@noop {} {\bibfield  {journal} {\bibinfo  {journal} {J. Phys. Commun.}\ }\textbf {\bibinfo {volume} {3}},\ \bibinfo {pages} {105005} (\bibinfo {year} {2019})}\BibitemShut {NoStop}%
  \bibitem [{\citenamefont {Wu}\ and\ \citenamefont {Shaknov}(1950)}]{Wu1950}%
  \BibitemOpen
  \bibfield  {author} {\bibinfo {author} {\bibfnamefont {C.~S.}\ \bibnamefont {Wu}}\ and\ \bibinfo {author} {\bibfnamefont {I.}~\bibnamefont {Shaknov}},\ }\href {\doibase 10.1103/PhysRev.77.136} {\bibfield  {journal} {\bibinfo  {journal} {Phys. Rev.}\ }\textbf {\bibinfo {volume} {77}},\ \bibinfo {pages} {136} (\bibinfo {year} {1950})}\BibitemShut {NoStop}%
\bibitem [{\citenamefont {Langhoff}(1960)}]{Langhoff1960}%
  \BibitemOpen
  \bibfield  {author} {\bibinfo {author} {\bibfnamefont {H.}~\bibnamefont {Langhoff}},\ }\href@noop {} {\bibfield  {journal} {\bibinfo  {journal} {Zeitschrift fur Physik}\ }\textbf {\bibinfo {volume} {160}},\ \bibinfo {pages} {186} (\bibinfo {year} {1960})}\BibitemShut {NoStop}%
\bibitem [{\citenamefont {Kasday}(1971)}]{Kasday1971}%
  \BibitemOpen
  \bibfield  {author} {\bibinfo {author} {\bibfnamefont {L.~R.}\ \bibnamefont {Kasday}},\ }in\ \href@noop {} {\emph {\bibinfo {booktitle} {Foundations of Quantum Mechanics: Proceedings of the International School of Physics "Enrico Fermi"}}},\ \bibinfo {editor} {edited by\ \bibinfo {editor} {\bibfnamefont {B.}~\bibnamefont {d'Espagnat}}}\ (\bibinfo  {publisher} {Academic P XIV, 1971},\ \bibinfo {year} {1971})\ pp.\ \bibinfo {pages} {195--210}\BibitemShut {NoStop}%
\bibitem [{\citenamefont {Faraci}\ \emph {et~al.}(1974)\citenamefont {Faraci}, \citenamefont {Gutkowski}, \citenamefont {Notarrigo},\ and\ \citenamefont {Pennisi}}]{Faraci1974}%
  \BibitemOpen
  \bibfield  {author} {\bibinfo {author} {\bibfnamefont {G.}~\bibnamefont {Faraci}}, \bibinfo {author} {\bibfnamefont {D.}~\bibnamefont {Gutkowski}}, \bibinfo {author} {\bibfnamefont {S.}~\bibnamefont {Notarrigo}}, \ and\ \bibinfo {author} {\bibfnamefont {A.~R.}\ \bibnamefont {Pennisi}},\ }\href {\doibase 10.1007/BF02763124} {\bibfield  {journal} {\bibinfo  {journal} {Lett. Nuovo Cimento}\ }\textbf {\bibinfo {volume} {9}},\ \bibinfo {pages} {607} (\bibinfo {year} {1974})}\BibitemShut {NoStop}%
\bibitem [{\citenamefont {Kasday}\ \emph {et~al.}(1975)\citenamefont {Kasday}, \citenamefont {Ullman},\ and\ \citenamefont {Wu}}]{Kasday1975}%
  \BibitemOpen
  \bibfield  {author} {\bibinfo {author} {\bibfnamefont {L.~R.}\ \bibnamefont {Kasday}}, \bibinfo {author} {\bibfnamefont {J.~D.}\ \bibnamefont {Ullman}}, \ and\ \bibinfo {author} {\bibfnamefont {C.~S.}\ \bibnamefont {Wu}},\ }\href@noop {} {\bibfield  {journal} {\bibinfo  {journal} {Il Nuovo Cimento B}\ }\textbf {\bibinfo {volume} {25}},\ \bibinfo {pages} {633} (\bibinfo {year} {1975})}\BibitemShut {NoStop}%
\bibitem [{\citenamefont {Wilson}\ \emph {et~al.}(1976)\citenamefont {Wilson}, \citenamefont {Lowe},\ and\ \citenamefont {Butt}}]{Wilson1976}%
  \BibitemOpen
  \bibfield  {author} {\bibinfo {author} {\bibfnamefont {A.}~\bibnamefont {Wilson}}, \bibinfo {author} {\bibfnamefont {J.}~\bibnamefont {Lowe}}, \ and\ \bibinfo {author} {\bibfnamefont {D.}~\bibnamefont {Butt}},\ }\href@noop {} {\bibfield  {journal} {\bibinfo  {journal} {J. Phys. G: Nucl. Phys.}\ }\textbf {\bibinfo {volume} {2}},\ \bibinfo {pages} {613} (\bibinfo {year} {1976})}\BibitemShut {NoStop}%
\bibitem [{\citenamefont {Bruno}\ \emph {et~al.}(1976)\citenamefont {Bruno}, \citenamefont {D'Agostino},\ and\ \citenamefont {Maroni}}]{Bruno77}%
  \BibitemOpen
  \bibfield  {author} {\bibinfo {author} {\bibfnamefont {M.}~\bibnamefont {Bruno}}, \bibinfo {author} {\bibfnamefont {M.}~\bibnamefont {D'Agostino}}, \ and\ \bibinfo {author} {\bibfnamefont {C.}~\bibnamefont {Maroni}},\ }\href@noop {} {\bibfield  {journal} {\bibinfo  {journal} {Nuovo Cim. B}\ }\textbf {\bibinfo {volume} {40}},\ \bibinfo {pages} {143} (\bibinfo {year} {1976})}\BibitemShut {NoStop}%
\bibitem [{\citenamefont {Bertolini}\ \emph {et~al.}(1981)\citenamefont {Bertolini}, \citenamefont {Diana},\ and\ \citenamefont {Scotti}}]{Bertolini1981}%
  \BibitemOpen
  \bibfield  {author} {\bibinfo {author} {\bibfnamefont {G.}~\bibnamefont {Bertolini}}, \bibinfo {author} {\bibfnamefont {E.}~\bibnamefont {Diana}}, \ and\ \bibinfo {author} {\bibfnamefont {A.}~\bibnamefont {Scotti}},\ }\href {\doibase 10.1007/BF02755105} {\bibfield  {journal} {\bibinfo  {journal} {Nuovo Cimento B}\ }\textbf {\bibinfo {volume} {63}},\ \bibinfo {pages} {651} (\bibinfo {year} {1981})}\BibitemShut {NoStop}%
\bibitem [{\citenamefont {Agostinelli}\ \emph {et~al.}(2003)\citenamefont {Agostinelli} \emph {et~al.}}]{Agostinelli2003}%
  \BibitemOpen
  \bibfield  {author} {\bibinfo {author} {\bibfnamefont {S.}~\bibnamefont {Agostinelli}} \emph {et~al.},\ }\href {\doibase 10.1016/S0168-9002(03)01368-8} {\bibfield  {journal} {\bibinfo  {journal} {Nucl. Instrum. Methods in Phys. Res. A}\ }\textbf {\bibinfo {volume} {506}},\ \bibinfo {pages} {250} (\bibinfo {year} {2003})}\BibitemShut {NoStop}%
\bibitem [{\citenamefont {Allison}\ \emph {et~al.}(2016)\citenamefont {Allison} \emph {et~al.}}]{Allison2016}%
  \BibitemOpen
  \bibfield  {author} {\bibinfo {author} {\bibfnamefont {J.}~\bibnamefont {Allison}} \emph {et~al.},\ }\href@noop {} {\bibfield  {journal} {\bibinfo  {journal} {Nucl. Instrum. Methods in Phys. Res. A}\ }\textbf {\bibinfo {volume} {835}},\ \bibinfo {pages} {186} (\bibinfo {year} {2016})}\BibitemShut {NoStop}%
\bibitem [{\citenamefont {Makek}\ \emph {et~al.}(2019)\citenamefont {Makek}, \citenamefont {Bosnar},\ and\ \citenamefont {Paveli{\'{c}}}}]{Makek2019a}%
  \BibitemOpen
  \bibfield  {author} {\bibinfo {author} {\bibfnamefont {M.}~\bibnamefont {Makek}}, \bibinfo {author} {\bibfnamefont {D.}~\bibnamefont {Bosnar}}, \ and\ \bibinfo {author} {\bibfnamefont {L.}~\bibnamefont {Paveli{\'{c}}}},\ }\href {\doibase 10.3390/condmat4010024} {\bibfield  {journal} {\bibinfo  {journal} {Condens. Matter}\ }\textbf {\bibinfo {volume} {4}},\ \bibinfo {pages} {24} (\bibinfo {year} {2019})}\BibitemShut {NoStop}%
\bibitem [{\citenamefont {Makek}\ \emph {et~al.}(2020)\citenamefont {Makek}, \citenamefont {Bosnar}, \citenamefont {Paveli{\'{c}}}, \citenamefont {{\v{S}}enjug},\ and\ \citenamefont {{\v{Z}}ugec}}]{Makek2020}%
  \BibitemOpen
  \bibfield  {author} {\bibinfo {author} {\bibfnamefont {M.}~\bibnamefont {Makek}}, \bibinfo {author} {\bibfnamefont {D.}~\bibnamefont {Bosnar}}, \bibinfo {author} {\bibfnamefont {L.}~\bibnamefont {Paveli{\'{c}}}}, \bibinfo {author} {\bibfnamefont {P.}~\bibnamefont {{\v{S}}enjug}}, \ and\ \bibinfo {author} {\bibfnamefont {P.}~\bibnamefont {{\v{Z}}ugec}},\ }\href {\doibase 10.1016/j.nima.2019.162835} {\bibfield  {journal} {\bibinfo  {journal} {Nucl. Instrum. Methods in Phys. Res. A}\ }\textbf {\bibinfo {volume} {958}},\ \bibinfo {pages} {162835} (\bibinfo {year} {2020})}\BibitemShut {NoStop}%
\bibitem [{\citenamefont {Newton}(2022)}]{RuthThesis}%
  \BibitemOpen
  \bibfield  {author} {\bibinfo {author} {\bibfnamefont {R.~N.}\ \bibnamefont {Newton}},\ }\href@noop {} {\bibfield  {journal} {\bibinfo  {journal} {PhD. thesis, University of York, UK}\ } (\bibinfo {year} {2022})}\BibitemShut {NoStop}%
\bibitem [{\citenamefont {Abdurashitov}\ \emph {et~al.}(2022)\citenamefont {Abdurashitov}, \citenamefont {Baranov}, \citenamefont {Borisenko}, \citenamefont {Guber}, \citenamefont {Ivashkin}, \citenamefont {Morozov}, \citenamefont {Musin}, \citenamefont {Strizhak}, \citenamefont {Tkachev}, \citenamefont {Volkov},\ and\ \citenamefont {Zhuikov}}]{Abdurashitov2022}%
  \BibitemOpen
  \bibfield  {author} {\bibinfo {author} {\bibfnamefont {D.}~\bibnamefont {Abdurashitov}}, \bibinfo {author} {\bibfnamefont {A.}~\bibnamefont {Baranov}}, \bibinfo {author} {\bibfnamefont {D.}~\bibnamefont {Borisenko}}, \bibinfo {author} {\bibfnamefont {F.}~\bibnamefont {Guber}}, \bibinfo {author} {\bibfnamefont {A.}~\bibnamefont {Ivashkin}}, \bibinfo {author} {\bibfnamefont {S.}~\bibnamefont {Morozov}}, \bibinfo {author} {\bibfnamefont {S.}~\bibnamefont {Musin}}, \bibinfo {author} {\bibfnamefont {A.}~\bibnamefont {Strizhak}}, \bibinfo {author} {\bibfnamefont {I.}~\bibnamefont {Tkachev}}, \bibinfo {author} {\bibfnamefont {V.}~\bibnamefont {Volkov}}, \ and\ \bibinfo {author} {\bibfnamefont {B.}~\bibnamefont {Zhuikov}},\ }\href {\doibase 10.1088/1748-0221/17/03/p03010} {\bibfield  {journal} {\bibinfo  {journal} {J. Instrum.}\ }\textbf {\bibinfo {volume} {17}},\ \bibinfo {pages} {P03010} (\bibinfo {year} {2022})}\BibitemShut {NoStop}%
\bibitem [{\citenamefont {Ivashkin}\ \emph {et~al.}(2023)\citenamefont {Ivashkin}, \citenamefont {Abdurashitov}, \citenamefont {Baranov}, \citenamefont {Guber}, \citenamefont {Morozov}, \citenamefont {Musin}, \citenamefont {Strizhak},\ and\ \citenamefont {Tkachev}}]{Ivashkin2023}%
  \BibitemOpen
  \bibfield  {author} {\bibinfo {author} {\bibfnamefont {A.}~\bibnamefont {Ivashkin}}, \bibinfo {author} {\bibfnamefont {D.}~\bibnamefont {Abdurashitov}}, \bibinfo {author} {\bibfnamefont {A.}~\bibnamefont {Baranov}}, \bibinfo {author} {\bibfnamefont {F.}~\bibnamefont {Guber}}, \bibinfo {author} {\bibfnamefont {S.}~\bibnamefont {Morozov}}, \bibinfo {author} {\bibfnamefont {S.}~\bibnamefont {Musin}}, \bibinfo {author} {\bibfnamefont {A.}~\bibnamefont {Strizhak}}, \ and\ \bibinfo {author} {\bibfnamefont {I.}~\bibnamefont {Tkachev}},\ }\href {\doibase 10.1038/s41598-023-34767-8} {\bibfield  {journal} {\bibinfo  {journal} {Sci. Rep.}\ }\textbf {\bibinfo {volume} {13}},\ \bibinfo {pages} {7559} (\bibinfo {year} {2023})}\BibitemShut {NoStop}%
\bibitem [{\citenamefont {Parashari}\ \emph {et~al.}(2023)\citenamefont {Parashari}, \citenamefont {Bosnar}, \citenamefont {Friščić}, \citenamefont {Kuncic},\ and\ \citenamefont {Makek}}]{parashari2023closing}%
  \BibitemOpen
  \bibfield  {author} {\bibinfo {author} {\bibfnamefont {S.}~\bibnamefont {Parashari}}, \bibinfo {author} {\bibfnamefont {D.}~\bibnamefont {Bosnar}}, \bibinfo {author} {\bibfnamefont {I.}~\bibnamefont {Friščić}}, \bibinfo {author} {\bibfnamefont {Z.}~\bibnamefont {Kuncic}}, \ and\ \bibinfo {author} {\bibfnamefont {M.}~\bibnamefont {Makek}},\ } (\bibinfo {year} {2023}),\ \Eprint {http://arxiv.org/abs/2304.11362} {arXiv:2304.11362 [quant-ph]} \BibitemShut {NoStop}%
  \bibitem [{\citenamefont {Caradonna}\ \emph {et~al.}(2024)\citenamefont {Caradonna}, \citenamefont {D'Amico}, \citenamefont {Jenkins},\ and\ \citenamefont {Watts}}]{Caradonna2024}%
  \BibitemOpen
  \bibfield  {author} {\bibinfo {author} {\bibfnamefont {P.}~\bibnamefont {Caradonna}}, \bibinfo {author} {\bibfnamefont {I.}~\bibnamefont {D'Amico}}, \bibinfo {author} {\bibfnamefont {D.~G.}\ \bibnamefont {Jenkins}}, \ and\ \bibinfo {author} {\bibfnamefont {D.~P.}\ \bibnamefont {Watts}},\ }\href {\doibase 10.1103/PhysRevA.109.033719} {\bibfield  {journal} {\bibinfo  {journal} {Phys. Rev. A}\ }\textbf {\bibinfo {volume} {109}},\ 033719 (\bibinfo {year} {2024})}\BibitemShut {NoStop}%
\bibitem [{\citenamefont {Caradonna}(2023)}]{Caradonna2023}%
  \BibitemOpen
  \bibfield  {author} {\bibinfo {author} {\bibfnamefont {P.}~\bibnamefont {Caradonna}},\ } (\bibinfo {year} {2023}),\ \Eprint {http://arxiv.org/abs/2402.12972} {arXiv:2402.12972 [hep-th]} { }\BibitemShut {NoStop}%
  \bibitem [{Note1()}]{Note1}%
  \BibitemOpen
  \bibinfo {note} {The current development version of QEG4 (v11.2) requires a filter on event ordering such that the polarization vectors are correctly assigned for all events.}\BibitemShut {Stop}%
\bibitem [{\citenamefont {{Di Francesco}}\ \emph {et~al.}(2016)\citenamefont {{Di Francesco}}, \citenamefont {Bugalho}, \citenamefont {Oliveira}, \citenamefont {Pacher}, \citenamefont {Rivetti}, \citenamefont {Rolo}, \citenamefont {Silva}, \citenamefont {Silva},\ and\ \citenamefont {Varela}}]{DiFrancesco2016}%
  \BibitemOpen
  \bibfield  {author} {\bibinfo {author} {\bibfnamefont {A.}~\bibnamefont {{Di Francesco}}}, \bibinfo {author} {\bibfnamefont {R.}~\bibnamefont {Bugalho}}, \bibinfo {author} {\bibfnamefont {L.}~\bibnamefont {Oliveira}}, \bibinfo {author} {\bibfnamefont {L.}~\bibnamefont {Pacher}}, \bibinfo {author} {\bibfnamefont {A.}~\bibnamefont {Rivetti}}, \bibinfo {author} {\bibfnamefont {M.}~\bibnamefont {Rolo}}, \bibinfo {author} {\bibfnamefont {J.~C.}\ \bibnamefont {Silva}}, \bibinfo {author} {\bibfnamefont {R.}~\bibnamefont {Silva}}, \ and\ \bibinfo {author} {\bibfnamefont {J.}~\bibnamefont {Varela}},\ }\href@noop {} {\bibfield  {journal} {\bibinfo  {journal} {J. Inst.}\ }\textbf {\bibinfo {volume} {11}} (\bibinfo {year} {2016})}\BibitemShut {NoStop}%
\end{thebibliography}
\end{document}


\preprint{APS/123-QED}

\title{Supplemental Materials: First Detailed Study of the Quantum Decoherence of Entangled Gamma Photons}
\maketitle
\section{Acceptance corrections from event mixing}

The acceptance correction applied to the data is

$$N_{corr}(\theta,\Delta\phi)=\frac{N_{real}(\theta,\Delta\phi)}{N_{mixed}(\theta,\Delta\phi)} \cdot \sum_{\theta,\Delta\phi=-\pi}^{\pi}N_{mixed}(\theta,\Delta\phi)$$
where $N_{real}(\theta,\Delta\phi)$ is the number of real coincident events within a given $\theta$ and $\Delta\phi$ bin, and $N_{mixed}(\theta,\Delta\phi)$ is the number of mixed events for the corresponding bin.

\section{Details of the QEG4-ENT model for entangled TCS}

\textcolor{black}{The QEG4-ENT model employs} the formalism of Snyder, Pasternack and Hornbostel (SPH) [17] (Eqn.~\ref{SNH}), derived in a ``partial polarization" ansatz in which the CS (described by Klein-Nishina theory) is factored from the Bell state. The DCS cross section in SPH formalism (using $\gamma$ nomenclature in Fig. 1) is given by: 

\begin{widetext}
\begin{equation}
     \frac{d^{2}\sigma}{d\Omega_{1}d\Omega_{2'}} = \frac{k_{1}^2k_{2'}^2(\alpha_{1}\alpha_{2'}-\alpha_{1}\sin^2(\theta_{2'})-\alpha_{2'}\sin^2(\theta_{1})+2\sin^2(\theta_{1})\sin^2(\theta_{2'})\sin^2(\Delta\phi))}{4\pi^2k_{1}^2(\frac{40}{9}-3\ln(3))^2}
     \label{SNH}
\end{equation}
\end{widetext}
where $\alpha_{1} = \frac{k_{1'}}{k_{1}} + \frac{k_{1}}{k_{1'}}$ and $\alpha_{2'} = \frac{k_{2''}}{k_{2'}} + \frac{k_{2'}}{k_{2''}}$ where $k_{1'}$ and $k_{2''}$ are the momenta of $\gamma_{1^\prime}$ and $\gamma_{2''}$. \textcolor{black}{For 511~keV $\gamma$ it reproduces the cross sections predicted in [16,17], but importantly also gives predictions for non-identical $\gamma$ energies.}
The DCS cross section (Eqn.~\ref{SNH}) is calculated for back-to-back photons. As in TCS $\gamma_{2}^{\prime}$ is no longer colinear with $\gamma_{1}$, the $\phi$ correlations are sampled taking $\hat{z}$ in the direction of $\gamma_{2}^{\prime}$ and $\phi$ relative to its polarization $\hat{\epsilon}_{\gamma_{2}^{\prime}}$ 
The orientation of $\hat{\epsilon}_{\gamma_{2}^{\prime}}$ is obtained from transforming $\hat{\epsilon}_{\gamma_{2}}$ into the frame of $\gamma_{2}^{\prime}$ without rotation, i.e. perpendicular to $\hat{\gamma}_{2}^{\prime}$ and in the plane formed by $\hat{\gamma}_{2}^{\prime}$ and the polarization vector of the incoming $\gamma$ ($\hat{\epsilon}_{\gamma_{2}}$).
Subsequent processes are simulated as separable $\gamma$.